\documentclass[pdflatex,referee,sn-mathphys-num]{sn-jnl}% Math and Physical Sciences Numbered Reference Style 
\usepackage{graphicx}%
\usepackage{multirow}%
\usepackage{amsmath,amssymb,amsfonts}%
\usepackage{amsthm}%
\usepackage{mathrsfs}%
\usepackage[title]{appendix}%
\usepackage{xcolor}%
\usepackage{textcomp}%
\usepackage{manyfoot}%
\usepackage{booktabs}%
\usepackage{algorithm}%
\usepackage{algorithmicx}%
\usepackage{algpseudocode}%
\usepackage{listings}%
\usepackage{pgf-pie}%
\usepackage{caption}
\usepackage{amsthm}
\usepackage{comment}
\usepackage{float}

\begin{document}

%\maketitle

\title[Article Title]{Physical Analogue Kolmogorov-Arnold Networks based on Reconfigurable Nonlinear-Processing Units}

\author[1]{\fnm{Manuel} \sur{Escudero}}%\email{iauthor@gmail.com}
\author[1]{\fnm{Mohamadreza} \sur{Zolfagharinejad}}
\author[2]{\fnm{Sjoerd} \sur{van den Belt}}%\email{iiauthor@gmail.com}
\author[2]{\fnm{Nikolaos} \sur{Alachiotis}}
\author*[1,3]{\fnm{Wilfred G.} \sur{van der Wiel}} \email{w.g.vanderwiel@utwente.nl}

\affil[1]{\orgdiv{NanoElectronics Group}, \orgname{MESA+ Institute and BRAINS Center for Brain-Inspired Computing, University of Twente}, \orgaddress{\street{P.O. Box 217}, \city{Enschede}, \postcode{7500 AE}, \country{The Netherlands}}}
\affil[2]{\orgdiv{CAES Group} and \orgname{BRAINS Center for Brain-Inspired Computing, University of Twente}, \city{Enschede}, \postcode{7500 AE}, \country{The Netherlands}}
\affil[3]{\orgdiv{Institute of Physics}, \orgname{University of M\"{u}nster}, \city{M\"{u}nster}, \postcode{48149} \country{Germany}}

\abstract{
Kolmogorov-Arnold Networks (KANs) shift neural computation from linear layers to learnable nonlinear edge functions, but implementing these nonlinearities efficiently in hardware remains an open challenge. Here we introduce a physical analogue KAN architecture in which edge functions are realized \textit{in materia} using reconfigurable nonlinear-processing units (RNPUs): multi-terminal nanoscale silicon devices whose input–output characteristics are tuned via control voltages. By combining multiple RNPUs into edge processors and assembling them into a reconfigurable analogue KAN (aKAN) with integrated mixed-signal interfacing, we demonstrate compact KAN-style regression and classification with programmable nonlinear transformations. Using experimentally calibrated RNPU models and hardware measurements, we demonstrate accurate function approximation across increasing task complexity while requiring fewer or comparable trainable parameters than multilayer perceptrons (MLPs). System-level estimates indicate an energy per inference of  roughly 200 pJ and an end-to-end inference latency of roughly 0.6 $\mu$s for a representative workload, corresponding to over 100$\times$ reduction in energy accompanied by $>$10$\times$ reduction in area compared to a digital fixed-point MLP at similar approximation error. These results establish RNPUs as scalable, hardware-native nonlinear computing primitives and identify analogue KAN architectures as a realistic silicon-based pathway toward energy-, latency-, and footprint-efficient analogue neural-network hardware, particularly for edge inference.
}

%\keywords{In-materia computing, Reconfigurable Nonlinear Processing Unit (RNPU), Energy-efficient neural-network hardware accelerator} 

\maketitle

\begin{figure}[tp!]
	\centering
     \makebox[\textwidth][c]{
     \includegraphics[width=18cm]{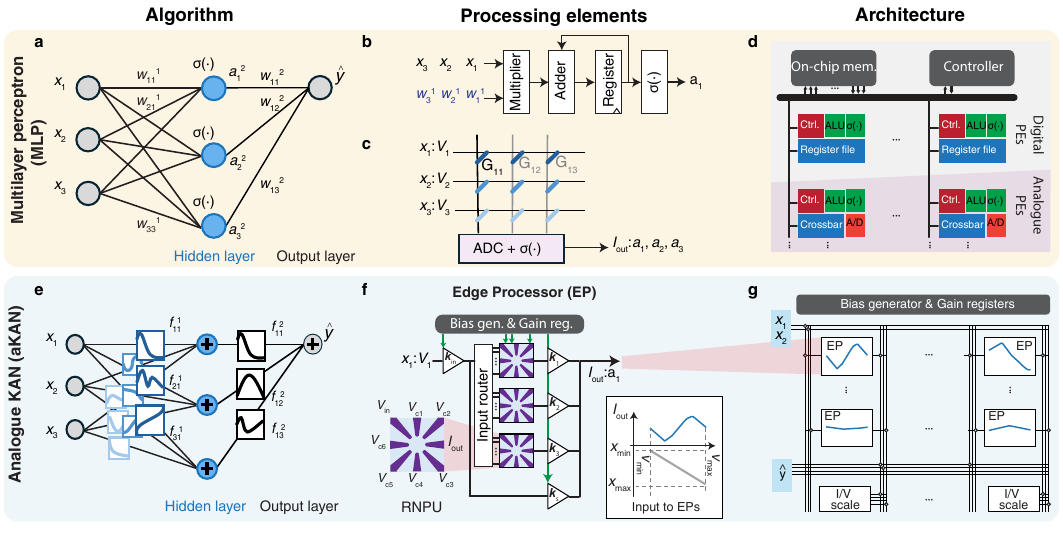}
     }
     \captionsetup{justification=justified}
     \caption{\textbf{Hardware implementations of neural networks and analogue Kolmogorov-Arnold networks.} \textbf{a}, Example of a multilayer perceptron (MLP) with three input neurons, a hidden layer and a single output neuron. Hidden neuron $j$ computes $a_j=\sigma\!\left(\sum_k w^{(1)}_{jk}x_k+b_j\right)$, where $w^{(1)}_{jk}$ denotes the weight from input neuron $k$ to hidden neuron $j$, $x_k$ the $k$-th input, $b_j$ an optional bias term, and $\sigma(\cdot)$ a fixed nonlinear activation function. The network output is $\hat{y}=\sum_j w^{(2)}_{1j}a_j$.
     \textbf{b}, Minimal block diagram of a digital circuit implementing a neuron. The dot product $\sum_k w^{(1)}_{jk}x_k$ is computed using multiply-accumulate (MAC) operations, followed by $\sigma(\cdot)$ to produce the activation $a_j$.
     \textbf{c}, analogue in-memory computing (AIMC) implementation of an MLP layer using a memristive crossbar array. Inputs ($x_j$) are mapped to input voltages $V_k$ that are fed to memristive devices with conductances $G_{jk}$, generating output currents $I_j=\sum_k G_{jk}V_k$, which are digitized by an analogue-to-digital converter (ADC) and passed through a digital activation function $\sigma(\cdot)$ to produce activations $a_j$.
    \textbf{d}, Generic hardware architecture for accelerating neural-network operations, organizing multiple digital and/or analogue processing elements (PEs) with local control logic (Ctrl.), arithmetic logic units (ALUs), register files, and, for analogue PEs, crossbar arrays and ADCs. The PEs interface with an on-chip memory unit and are coordinated by a controller to parallelize MAC operations. 
    \textbf{e}, Schematic of a Kolmogorov-Arnold Network (KAN) [14], with learnable nonlinear edge functions $f^{l}_{ij}(\cdot)$ replacing fixed node activation functions, and node operations primarily consisting of summing incoming edge outputs.
    \textbf{f}, Edge processor (EP) implementing KAN edge functions using parallel reconfigurable nonlinear-processing units (RNPUs). Each RNPU is configured by selecting an input electrode (via an input router) and tuning its nonlinear transfer characteristic using control voltages ($V_{c,i}$); output gains and optional skip connections further increase edge-function flexibility. The EP transforms the linearly encoded input $x_i$ as a voltage $V_i$ within the RNPU operating range to $I_{out}$, which is the sum of the scaled RNPU outputs (inset). 
    \textbf{g}, analogue KAN (aKAN) architecture comprising a reconfigurable array of EPs interconnected via programmable switch matrices. Bias generator and gain registers configure EP characteristics. Intermediate I/V scaling blocks rescale signals for subsequent layers. A final linear readout produces the prediction $\hat{y}$.}
     \label{fig:introduction}
\end{figure}

\section*{Introduction}

Multilayer perceptrons (MLPs) are the foundational building blocks of modern deep learning models, such as convolutional neural networks (CNNs) for machine vision \cite{Krizhevsky2017} and generative pre-trained transformers (GPTs) for natural language processing tasks \cite{Vaswani2017}. As illustrated in Fig. \ref{fig:introduction}a, an MLP consists of fully connected layers of neurons organized into an input layer, one or more hidden layers, and an output layer. Each layer applies an affine transformation followed by a nonlinear activation function, $\sigma(\cdot)$, such as the hyperbolic tangent (tanh) or the rectified linear unit (ReLU). Based on the universal approximation theorem, MLPs can approximate any function to an arbitrary degree of accuracy under suitable conditions  \cite{MLP_theorem}. The computations performed by an MLP can be decomposed into linear operations, i.e., multiply–accumulate (MAC) operations, and nonlinear operations, i.e., activation functions. In conventional digital hardware, this corresponds to first computing dot products as a sequence of MAC operations (Fig.~\ref{fig:introduction}b), followed by evaluation of the nonlinear activation function. Although the linear part accounts for the vast majority of operations in deep neural networks, nonlinear activations are essential for expressivity and learning capacity \cite{activation_function}. 

Massive computational resources have driven the rise of deep neural networks, supported by parallel accelerators such as graphics processing units and, more recently, application-specific integrated circuits and analogue in-memory computing (AIMC) platforms based on memristive crossbar arrays (Fig. \ref{fig:introduction}c) \cite{Markovic2020, Reza2024, Singh2025}. These platforms have enabled both the training of larger models and their deployment for on-device and edge inference, typically by organizing multiple digital and/or analogue processing elements (PEs) around on-chip memory and a central controller to parallelize MAC operations (Fig.~\ref{fig:introduction}d). Crucially, both the theory and practice of modern deep learning are dominated by linear algebra at scale: most floating-point operations arise from matrix–vector and matrix–matrix multiplications, which are naturally mapped onto parallel hardware. As a result, today's hardware ecosystem is optimized primarily for linear operations, reflecting the tight coupling between algorithmic progress and prevailing compute substrates \cite{Sutton2019, Hooker2021, Laydevant2024}. By contrast, \textit{nonlinear} operations are critical for representational power and effective learning, yet more difficult to implement in conventional digital and mixed-signal circuits. Therefore, they have received far less architectural specialization. 

Although linear operations in MLPs can be effectively accelerated on specialized hardware, their computational cost is increasing in a non-sustainable way \cite{Thompson2020, Mehonic2022}, limiting deployment in resource-constrained environments. Moreover, their limited interpretability renders them black boxes, hindering scientific discovery and trustworthy deployment in domains such as healthcare, finance, and security \cite{Zhang2021}. Kolmogorov-Arnold Networks (KANs) \cite{KAN} have recently emerged as a promising alternative by shifting the representational emphasis from large linear transformations followed by fixed activation functions to learnable univariate nonlinear functions placed on the edges. Rooted in the Kolmogorov–Arnold representation theorem \cite{kan_theorem}, KANs express multivariate continuous functions as structured compositions and sums of learned univariate functions, as illustrated in Fig.~\ref{fig:introduction}e. When combined with modern training techniques such as backpropagation, KAN variants have demonstrated regression performance comparable to and in some cases, exceeding that of MLP baselines while often using substantially fewer parameters \cite{KAN, Yu2024}. 

This shift in representation has important hardware implications. Whereas current AI hardware is optimized for linear, matrix-style MAC computations, KANs - in contrast to MLPs - devote a significant fraction of their computation to nonlinear operations such as univariate spline evaluations. As a result, reduced model size does not necessarily translate into higher energy efficiency, motivating a careful energy analysis. Recent studies have proposed hardware acceleration of KAN models, including digital implementations using look-up tables (LUTs) \cite{KAN_digital_hardware_LUTs, Hoang2025} and compute-in-memory architectures that map KAN edge functions to piecewise-linear segments evaluated directly in hardware \cite{Sudarshan2025}. Beyond purely digital acceleration, photonic approaches have also been proposed, mapping KAN edge nonlinearities to integrated optical building blocks such as Mach-Zehnder interferometers and ring-assisted MZI units \cite{Sozos2026, PhotonicKAN}. More recently, a fully analogue KAN concept based on negative-differential-resistance tunnel-diode characteristics has been introduced and evaluated using simulated device I-V curves combined with surrogate/spline fitting \cite{Li2025}. These developments motivate a complementary direction: implementing KAN nonlinearities via physical computing, i.e., leveraging the intrinsic properties of matter to compute directly in the physical domain. We have articulated such perspectives in the context of “intelligent matter” \cite{Kaspar2021} and computation with physical dynamical systems \cite{Jaeger2023}. See also \cite{Liang2024, Kasai2022} for reviews of physical reservoir computing with emerging electronics and Momeni \textit{et al.} for a review of training strategies for physical neural networks, spanning both simulation-based (digital-twin) workflows and experimental demonstrations across physical platforms \cite{Momeni2025}. From this viewpoint, a promising route toward efficient KAN hardware is to develop scalable physical computing primitives that realize trainable nonlinear edge functions natively through tunable device physics, rather than through LUT-style emulation or circuit-level approximations. As this manuscript was being completed, we became aware of the contemporaneous preprint by Taglietti \textit{et al.}~\cite{Taglietti2026}, which independently explores physical KAN implementations using reconfigurable nonlinear silicon devices, further underscoring the timeliness of hardware-native nonlinear edge functions. Distinctively, we develop a complementary RNPU-based analogue KAN platform in which each edge processor composes multiple RNPUs to realize richer nonlinear edge functions; we further demonstrate device-to-device robustness, pruning for compact deployment under resource constraints, and quantitative mixed-signal energy, latency, and area analyses, including peripheral circuitry, to identify system-level overheads and bottlenecks.

Here, we substantially extend our preliminary account \cite{RezaThesis2025} and demonstrate a \textit{physical} analogue KAN architecture using reconfigurable nonlinear-processing units (RNPUs) as physical computing primitives. Specifically, we implement the defining operation of KANs by replacing fixed activation functions with learnable nonlinear functions on the network edges, realized \textit{in materia} by RNPUs. Figure \ref{fig:introduction}f shows the RNPU concept: a multi-terminal nanoscale doped-silicon device whose steady-state input-output response can be continuously tuned by applying control voltages to its electrodes. In our earlier work, we pioneered the use of RNPUs as physical function approximators and demonstrated that a single device can be trained to realize a variety of time-independent computational tasks, including classification and regression in static settings at low temperature \cite{Chen2020, Chen2021, RuizEuler2020,RuizEuler2021, Ven2023, Boon2025}. More recently, we extended the concept of RNPUs to time-dependent signal processing tasks at room temperature, where RNPUs implement nonlinear temporal transformations that enable efficient processing of dynamic input streams \cite{Reza2025}. Our prior results thus establish RNPUs as a versatile platform for programmable nonlinear computation.
    
This makes KANs a natural architectural match: unlike MLPs, which rely primarily on linear layers combined with simple fixed activations, KANs place the computational emphasis on expressive trainable nonlinear edge functions. RNPUs directly provide such nonlinear primitives in hardware, enabling edge functions to be realized \textit{in materia} without reducing them to digital approximations or LUT-based emulations. In this study, we use each RNPU in a single-input, single-output configuration, thus effectively implementing an \textit{in-materia} nonlinear data processor, while the remaining electrodes are biased with control voltages that tune the nonlinear characteristics of the device. The RNPU is operated in steady-state mode (see Materials and Methods). To increase expressivity and precision, multiple RNPUs can be combined to form a reconfigurable edge processor (EP) whose aggregate response implements the desired learned nonlinearity (see Fig.~\ref{fig:introduction}f). Multiple EPs can then be combined to form an RNPU-based analogue KAN (aKAN) architecture, as schematically presented in Fig.~\ref{fig:introduction}g. The resulting aKAN architecture follows a design philosophy similar to that of a field-programmable analogue array \cite{George2016}: programmable interconnections provide flexible routing between EPs, enabling multi-layer, variable-width networks. In addition, intermediate I/V conversion and rescaling blocks ensure that EP output currents are mapped to voltage ranges compatible with subsequent EP stages. In the example shown in the lower part, each input variable $x_i$ is first linearly encoded as a voltage within the operating range of the RNPU $[V_{\min},V_{\max}]$ and applied to an EP, which acts as a nonlinear map to an output current $I_{\mathrm{out}}$. These EP outputs are then combined by the network to generate the final prediction $\hat{y}$, illustrated here for a two-input target function. 

We evaluate our aKAN architecture on regression and classification benchmarks using experimentally calibrated surrogate models together with experimental hardware measurements. For representative function-approximation tasks, aKANs reach $\mathrm{MSE}\sim10^{-2}$ (and below for selected cases) and approach the performance of software MLP baselines, including MLPs with tanh activation function, at comparable model sizes. Using measured RNPU characteristics ($\sim$50 nW power, $\sim$1 $\mu$m$^{2}$ footprint, $\sim$10 ns intrinsic response time \cite{Reza2025}) and realistic mixed-signal peripheral assumptions, our system-level hardware estimates indicate up to three orders of magnitude lower energy per inference than a digital fixed-point MLP at similar error levels, while reducing the estimated silicon area by roughly one order of magnitude at sub-micosecond latency. Although the quantitative system-level estimates are based on a representative device, we further show that aKAN functionality is robust to device-to-device variability by evaluating inference with multiple surrogate models calibrated to different physical RNPUs. These results position aKANs as a route toward compact, low-power edge inference where programmable nonlinear transformations are implemented directly in matter.

We employ a hybrid training and validation workflow that (a) trains the RNPUs to identify control voltages that give a desired nonlinear edge function, and (b) validates nonlinear functions by performing experimental measurements. Thus, we first develop data-driven surrogate models for six devices that allow for end-to-end training in software using backpropagation \cite{RuizEuler2021}. A surrogate model captures how the steady-state output current of an RNPU depends on the applied input voltage at a chosen input electrode and on the set of control voltages applied to the remaining six electrodes, which makes gradient-based optimization possible. We then embed the surrogate model into Brains-Py, an open-source material-learning framework \cite{AlegreIbarra2023}, which can include multiple RNPUs as well as the surrounding circuit-level degrees of freedom required by our architecture (for example, electrode selection via the input router, programmable output gains, and optional skip connections inside the edge processor).

\section*{Results}

\subsection*{Function approximation with analogue KANs}

\begin{figure}[tp!]
	\centering
     \makebox[\textwidth][c]{
     \includegraphics[width=18cm]{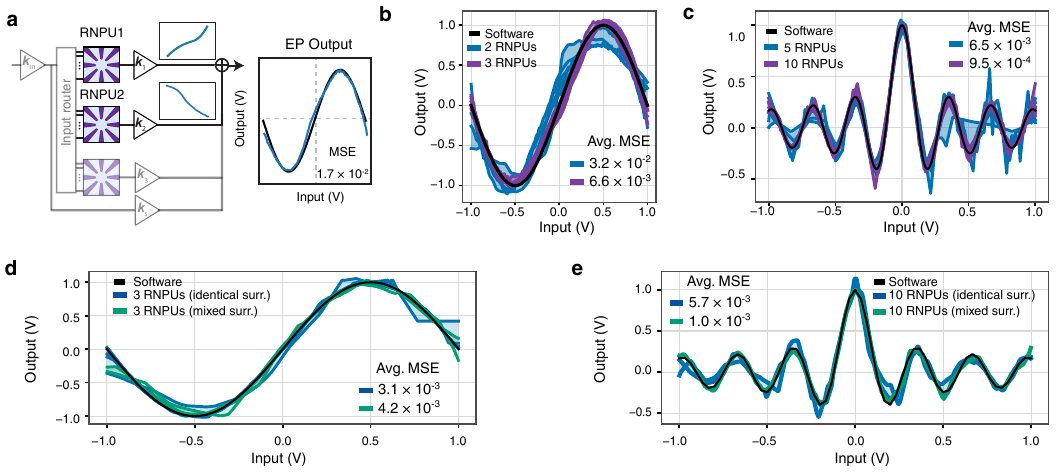}
     }
     \caption{\textbf{Function approximation with RNPU-based edge processors.} \textbf{a}, Experimental sine-wave fitting using an edge processor (EP) composed of two RNPUs in parallel. A linear combination of the individual RNPU outputs (left panels) approximates (blue curve) the target sine function (black curve). 
     \textbf{b},  Experimental sine-wave approximation using EPs composed of 2 (blue) and 3 (purple) RNPUs in parallel. The shaded regions indicate the spread among five fits obtained from independent training runs, using the best configurations identified from a prior random exploration of input-electrode selections.
     \textbf{c}, Bessel-function approximation ($\mathrm{J}_0(20x)$) using EPs with 5 (blue) and 10 (purple) parallel RNPUs. The shaded regions show the spread among five fits, as in panel \textbf{b}.
     \textbf{(d)--(e)}, Function approximation of \textbf{d} sine-wave and \textbf{e} Bessel function considering surrogate models based on up to 6 different RNPUs, in a single EP. Both homogeneous configurations (identical surrogates, in blue) and heterogeneous configurations (mixed surrogates, in green) for the EP are considered.
     }
     \label{fig:EP_regression}
\end{figure} 

We use standard backpropagation to optimize two coupled sets of parameters: (1) per-RNPU control voltages, which shape the nonlinear transfer characteristics and therefore implement the learned univariate edge nonlinearities, and (2) the linear combination coefficients (input and output gains) that determine how multiple RNPU responses are summed to form each edge function. After training, the same parameter set can be evaluated purely in software (using the surrogate model), or transferred to the physical device for experimental validation. Because our current measurement setup includes a single RNPU, multi-RNPU networks are realized via time multiplexing: each RNPU instance in the trained aKAN is emulated sequentially by applying its corresponding input and control voltages to the same physical device and recording the output current, while intermediate node values and summations are handled externally. In this way, we exploit \textit{in-materia} nonlinear computation at the edges, while the summation of edge outputs at the nodes and the bookkeeping needed to route and rescale signals across layers are still done conventionally.

We first demonstrate the RNPUs' capability to approximate continuous nonlinear functions, using the EP concept introduced in Fig.~\ref{fig:introduction}f. As a simple univariate example, we fit a sine wave by combining the outputs of multiple RNPUs with trainable output gains. This choice is deliberate: we find that a single RNPU does not provide sufficient functional complexity to accurately reproduce this target function. Figure~\ref{fig:EP_regression}a shows the resulting fit together with the individually learned RNPU responses, obtained by tuning the control voltages using our training procedure. For the case of two devices, we obtain a mean-squared error (MSE) of $1.7\times10^{-2}$, using 15 trainable parameters in total (six control-voltage parameters per RNPU, plus output gains and a single EP input gain when applicable).

Next, we investigate how the approximation accuracy scales with the number of RNPUs and varies depending on the selected input electrodes. Each RNPU operates using a single chosen input electrode, and different choices of input electrodes among the RNPUs in an EP can lead to markedly different performance. Increasing the number of RNPUs per EP, as well as varying the chosen input electrodes among RNPUs, improves expressivity and enables more complex function representations. We illustrate this by approximating both the sine wave and the Bessel function $\mathrm{J}_0(20x)$ using different numbers of RNPUs (see Figs.~\ref{fig:EP_regression}b and c, respectively). The Bessel function is a challenging and relevant (ubiquitous in wave and diffraction problems) oscillatory benchmark with slowly decaying amplitude, which probes the ability of the network to represent nontrivial nonlinear structure beyond simple periodic functions. For each configuration, we perform multiple training runs with randomized input electrodes and report the best five results; the shaded region indicates the spread among these solutions. In both cases, using more devices (\textit{e.g.}, three for the sine wave and ten for $\mathrm{J}_0(20x)$) systematically lowers the approximation error, consistent with the increased nonlinear expressivity afforded by larger RNPU ensembles. We use these two examples to confirm similar performance with all 6 RNPU devices considered in this work, both in a homogeneous (same surrogate model) and heterogeneous (multiple surrogate models) arrangements (Figs.~\ref{fig:EP_regression}d--e). These results indicate that device-to-device variability does not degrade approximation performance when multiple surrogate models are combined within a single EP.

\begin{figure}[tp!]
	\centering
     \makebox[\textwidth][c]{
     \includegraphics[width=18cm]{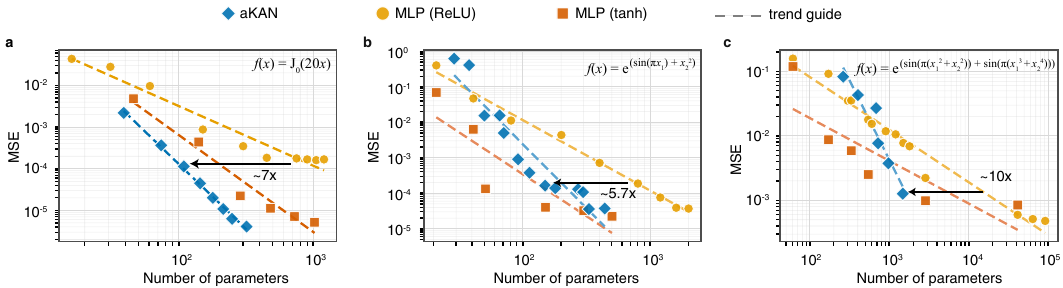}
     }
     \caption{\textbf{Benchmarking aKANs for regression tasks.} 
     \textbf{a--c}, Scalability of aKANs and MLP baselines (with ReLU or tanh activations), showing approximation error as a function of the number of trainable parameters for three target functions: $\mathrm{J}_0(20x)$ in panel \textbf{a} $e^{(\sin(\pi x_1)+x_2^2)}$ in panel \textbf{b} and $e^{(\sin(\pi(x_1^2+x_2^2)) + \sin(\pi(x_3^2+x_4^2)))}$ in panel \textbf{c}. Both aKANs and MLPs are simulated across a range of network depths and widths; for aKANs, we additionally vary the number of RNPUs per EP and randomly select RNPU input electrodes. The specific architectures used are listed in Supplementary Tables \ref{tab:MLP_conf} and \ref{tab:aKAN_conf}. Each panel includes a representative parameter reduction factor of the aKAN relative to an MLP with ReLU.
     }
     \label{fig:regression_scaling}
\end{figure}

Furthermore, we assess how the proposed aKAN architecture scales relative to conventional MLP baselines. To this end, we compare algorithmic performance, quantified as MSE, as a function of the number of trainable parameters (\textit{i.e.}, model size). We evaluate approximation performance on representative 1-, 2-, and 4-variable target functions adopted from Ref.~\cite{KAN}, which provide well-defined benchmarks with increasing functional complexity. For each target, we simulate both aKANs and MLPs across a range of depths and layer widths. For aKANs, we additionally sweep the number of RNPUs per edge function and randomly select the input electrodes for each RNPU. For the MLP baselines, we consider networks with either ReLU or tanh activation functions at the nodes, evaluated in software using 32-bit floating-point precision. It is worth noting that, while the results correspond to the reconfigurability of a single device, we also confirm a systematic reduction of the approximation error while evaluating the function in Fig.~\ref{fig:regression_scaling}b on up to four surrogate model simultaneously placed in the same EP (Fig. \ref{figS:mult_devices_regression}).

The resulting scaling trends are summarized in Figs.~\ref{fig:regression_scaling}a--c. 
For all three target functions, the approximation error systematically decreases with increasing model size (number of trainable parameters), for both aKANs and MLPs. When comparing error versus parameter count, the aKAN performance lies between that of ReLU- and tanh-MLPs, approaching the tanh baseline for sufficiently large aKANs. This behaviour is expected: the 2- and 4-variable functions require a minimum aKAN architecture to represent the target as compositions of univariate edge functions. Once this minimum size is reached, the aKAN can effectively exploit the expressivity of learned nonlinear edges, and its performance improves rapidly. In this regime, aKANs substantially outperform ReLU-based MLPs at comparable model sizes. Although tanh-MLPs remain competitive in this numerical comparison, it should be noted that tanh provides a highly expressive nonlinearity and is not straightforward to implement efficiently in hardware, particularly with the software baseline assumed here (32-bit floating-point arithmetic).

\subsection*{Classification with analogue KANs}
We next evaluate aKANs for two binary classification tasks: (i) synthetic two-dimensional tunable datasets and (ii) real-world benchmark datasets. Because these problems involve few input features and a low-dimensional output, compact network architectures are sufficient. In all cases, the model produces a single scalar output whose magnitude determines the assigned class.

\begin{figure}[tp!]
	\centering
     \makebox[\textwidth][c]{
     \includegraphics[width=18cm]{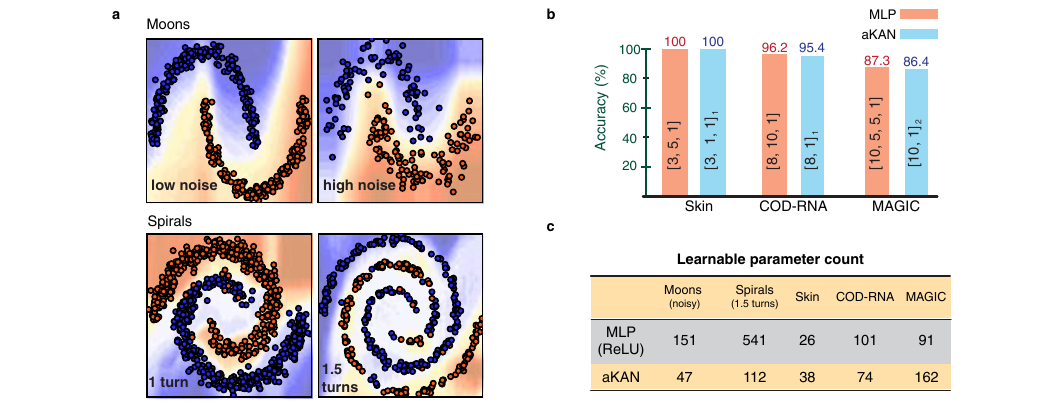}
     }
     \caption{\textbf{Benchmarking aKANs for binary classification.}
\textbf{a}, Experimental classification and decision boundaries on the Moons and Spirals datasets by aKANs. 
Top left: low perturbed (noise = 0.05) Moons, where a $[2,1]_2$ aKAN achieves 100\% accuracy, compared to 99.1\% for a $[2,9,9,1]$ MLP (not shown). Top right: highly perturbed (noise = 0.15) Moons, where a $[2,1]_3$ aKAN reaches 100\% accuracy versus 98.7\% for a $[2,10,10,1]$ MLP (not shown). Bottom left: 1-turn spiral, where a $[2,3,1]_4$ aKAN achieves 98.4\% accuracy compared to 98.5\% for a $[2,10,10,1]$ MLP (not shown). Bottom right: 1.5-turns spiral, where a $[2,4,1]_5$ aKAN achieves 91.3\% accuracy, matching 91\% for a $[2,15,15,15,1]$ MLP. 
Networks are denoted $[n_I,n_{H1},\ldots,n_{HL},n_O]_{d}$, with $n_I$, $n_O$ the input/output sizes, $n_{H1}\ldots n_{HL}$ the hidden layer widths, and $d$ the RNPU number per EP in aKANs. \textbf{b}, Simulated aKAN accuracy on Skin Segmentation, COD-RNA, and MAGIC datasets, compared with ReLU-MLPs; architectures indicated. \textbf{c}, Comparison of learnable parameter count for aKANs and ReLU-MLPs across the classification tasks, illustrating that aKANs can require up to $\sim5\times$ fewer parameters to reach comparable accuracy.
}

     \label{fig:classification}
\end{figure}

We first consider the Moons and Spirals datasets (Fig.~\ref{fig:classification}a) \cite{moons,spirals}, which provide controlled settings with progressively more nonlinear decision boundaries. For each dataset, we generate two variants of increasing complexity by varying a single generative parameter: the noise level for Moons (top panels) and the number of spiral turns for Spirals (bottom panels). Next, we experimentally explore a range of aKAN architectures, from minimal to larger configurations, and stop increasing complexity once accuracy gains saturate. Similarly, we evaluate ReLU-MLPs with minimal learnable parameters and hidden layers that can achieve comparable classification accuracies (within a $<1\%$-point difference). Figure \ref{fig:classification}a summarizes the aKAN performance under these conditions. As expected, the required model size increases with the geometric complexity of the decision boundary for both aKANs and MLPs.

We also evaluate the performance of aKANs in three benchmark datasets: MAGIC Gamma Telescope \cite{magic}, COD-RNA \cite{codrna}, and Skin Segmentation \cite{skin}. Our aim is not only to demonstrate feasibility, but also to determine where aKANs offer advantages over conventional neural networks. As before, we identify aKANs that balance architectural complexity and accuracy. For comparison, we pair each aKAN with a ReLU-MLP whose average accuracy differs by no more than 1\%-point from that of its aKAN counterpart. Simulations (Fig.~\ref{fig:classification}b) show that aKANs achieve software-iso accuracy, although the Skin and MAGIC datasets require a larger number of learnable parameters. Notably, here, each RNPU accounts for six control-voltage parameters plus the output gain for each EP. For classification tasks that do not require complex nonlinear transformations, this level of parametrization may be excessive, making the parameter count per RNPU suboptimal for the aKAN. Instead, we observe comparable accuracies with simpler architectures that have no hidden layer or with only a single RNPU per edge.. 

We compare parameter efficiency with ReLU-MLP baselines and summarize the corresponding parameter counts in Fig.~\ref{fig:classification}c. For geometric datasets, aKANs achieve competitive accuracy using compact architectures with substantially fewer trainable parameters, consistent with their ability to represent nonlinear decision boundaries through expressive learned edge functions. For real-world datasets, accurate classification is already obtained with very simple aKANs, likely because these benchmarks are largely separable through a small number of dominant features and low-order statistics, a regime well matched to the representational capacity of single RNPUs. Under such conditions, the parameter-efficiency advantage of larger aKANs is not fully revealed, and the architecture may effectively be over-parameterized for these tasks. Even when reducing the number of control parameters (e.g., by using a single RNPU per edge function and fewer control electrodes per device), the impact on classification accuracy is limited. However, from a system-level perspective, such parameter reductions do not substantially change overall efficiency because, as discussed below, system-level performance is currently dominated by mixed-signal overhead rather than by the number of nonlinear control parameters.

\subsection*{Sparsification strategy in analogue KANs}
In the previous examples, prior task knowledge enabled the straightforward selection of suitable aKAN architectures. However, in practice, this information is often not available, and MLPs are typically oversized to facilitate training convergence, at the expense of higher computational costs. In KANs, however, interpretable edge functions enable structural pruning that minimizes network size and is more efficient in deployment. In particular, L1 and entropy-based regularizations, which promote sparsity and interpretability, have been already discussed in the literature that enable the removal of edges with a negligible contribution \cite{KAN}.

\begin{figure}[t!]
	\centering
     \makebox[\textwidth][c]{
     \includegraphics[width=18cm]{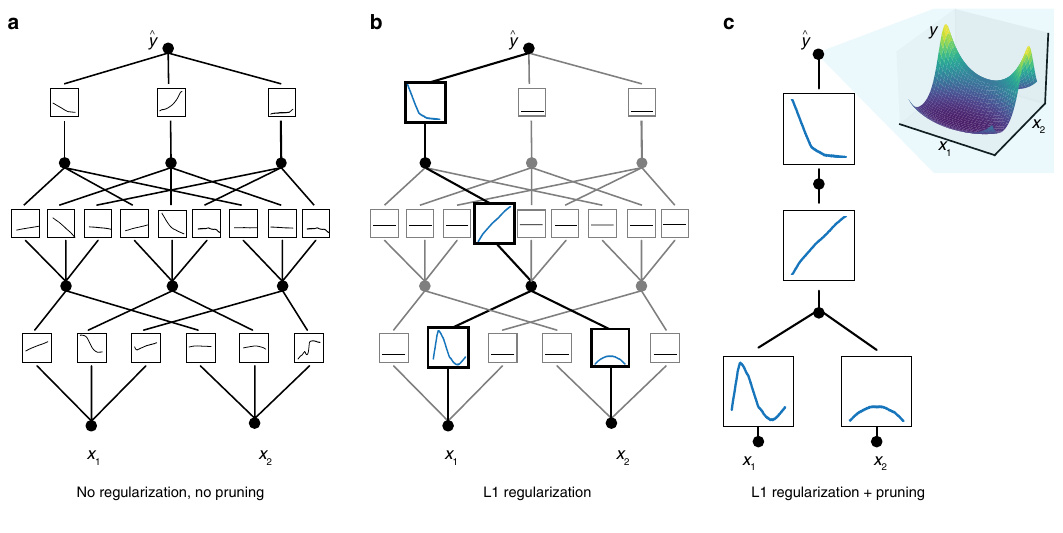}
     }
     \caption{\textbf{Regularization and pruning of analogue KANs.} \textbf{a}, aKAN $[2,3,3,1]_{3}$ fit to $y=e^{(\sin(\pi x_1)+x_2^2)}$ without regularization or pruning (MSE $=8\times10^{-3}$). \textbf{b}, Same task and aKAN with L1 regularization applied to RNPU output gains (MSE $=2.38\times10^{-2}$); the contributions of many EPs are suppressed and the dominant EPs are highlighted. \textbf{c}, Pruned network to $[2,1,1]_{3}$ derived from \textbf{b} after fine-tuning the remaining EPs (MSE $=2.33\times10^{-2}$).}

     \label{fig:pruning}
\end{figure}

In our proposed aKAN architecture, pruning yields more compact models and frees up EPs, thereby reducing hardware utilization. Thus, we investigate whether aKANs can be pruned without sacrificing performance. Following the node-pruning philosophy \cite{KAN}, we prune EPs using a simplified scheme: we add an output gain per EP and apply L1 regularization to these gains to suppress EP contributions to the next layer. After training, the regularization gains are absorbed into EP gains $k_i$. As a case study, we revisit the 2-variable function $e^{(\sin(\pi x_1)+x_2^2)}$ (Fig.~\ref{fig:regression_scaling}b). Although it admits a minimal aKAN[2,1,1]$_3$ representation, we start from an oversized aKAN[2,3,3,1]$_3$ and compare training without regularization versus L1 regularization applied to all EP output gains, see Figures ~\ref{fig:pruning}a and~\ref{fig:pruning}b, respectively.

L1 regularization promotes sparsity by driving many EP gains toward near-zero values, thus suppressing their contribution to the output. The result is a more interpretable model (Fig. \ref{fig:pruning}b): the remaining active EPs align with meaningful input-output dependencies of the target function. These are reflected by the learned transfer characteristics that resemble elementary components such as sine, square, and exponential operations.

For pruning, we remove an EP if (i) it is insufficiently activated by the preceding layer (below a threshold) or (ii) it contributes only weakly to the subsequent activation. Applying this rule to the regularized network in Fig. \ref{fig:pruning}b results in substantial pruning. After fine-tuning, the compact network (Fig. \ref{fig:pruning}c) retains the essential EP transfer characteristics needed for approximation. The second-layer EP becomes largely linear and effectively serves as a skip connection, consistent with the fact that the initial aKAN[2,3,3,1]$_{3}$ includes an additional layer relative to the minimal target composition. Although the regularized and pruned models achieve similar MSE, both remain less accurate than the unpruned network. This single case study shows that pruning can reduce hardware utilization while preserving the dominant functional structure, although there is a trade-off with approximation accuracy. 

\subsection*{System-level hardware comparison with digital neural networks}
Having established the regression and classification performance of aKANs, we now benchmark an RNPU-based aKAN against a conventional digital tanh-MLP at the system level in terms of energy per inference, silicon area, and latency. Because the two approaches rely on fundamentally different computational primitives and design architectures and our goal is not to evaluate a general-purpose implementation, we perform the comparison for a representative nonlinear workload: approximation of $e^{(\sin(\pi x_1) + x_2^2)}$  (Fig. \ref{fig:regression_scaling}b). We therefore adopt the aKAN and MLP configurations from Fig.~\ref{fig:regression_scaling}b and estimate the energy, area, and latency based on the hardware components required by these implementations.

We begin by characterizing the response of a single RNPU, which provides the physical basis for the system-level estimates. Applying representative input and control biases yields an average power of 50 nW. The footprint of an individual RNPU is approximately 1 $\mu$m$^2$, including the device area and routing overhead, and the intrinsic inference latency is $\sim$10 ns \cite{Reza2025}. 

\begin{figure}[t!]
	\centering
     \makebox[\textwidth][c]{
     \includegraphics[width=18.0cm]{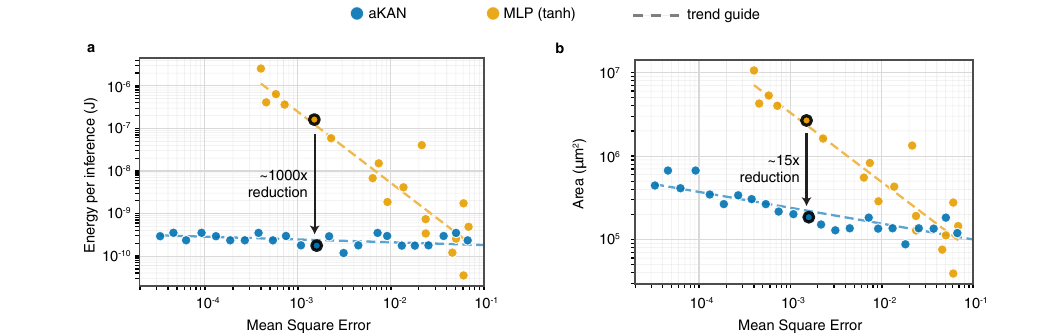}
     }
     \caption{\textbf{System-level energy and silicon area comparison between aKANs and MLPs for nonlinear function evaluation.} Both architectures are trained to approximate $e^{(\sin(\pi x_1)+x_2^2)}$, and the energy and area costs shown correspond to inference on the trained networks. For the digital baseline, MLP multiply-accumulate units are implemented using the NanGate45 Open Cell library and the tanh activation function is realized by a look-up table. \textbf{a}, Estimated energy per inference as a function of mean-squared error (MSE). The transimpedance amplifier (TIA) is the dominant contribution to the aKAN energy consumption. \textbf{b}, Estimated silicon area versus MSE. Multiple network widths and depths are included for both aKAN and MLP architectures. Solid red lines serve as guides to the eye. Two representative results (outlined dots) illustrate the reduction in energy and area of the aKAN approach.
     }
     \label{fig:hardware}
\end{figure}

In addition to the RNPUs, aKANs require peripheral circuitry to generate input and control voltages, acquire RNPU outputs, and interface with the surrounding system. This mixed-signal overhead, comprising digital-to-analogue converters (DACs), ADCs, and current-to-voltage conversion, contributes to the overall energy consumption, silicon area, and latency. We assume 10-bit DACs with a power consumption of 1.46 $\mu$W at 2 MHz, both for quasi-static control voltages and input ports \cite{DAC}, and a 12-bit, 100 MS/s ADC (2.6 mW) for output digitization \cite{ADC}. Programmable output gains are implemented using memristive elements and one transimpedance amplifier (TIA) per node. For this stage, we adopt an integrated low-power TIA design with a power consumption of 94 $\mu$W, which satisfies the required bandwidth and gain constraints  \cite{Mathew2022}. In practice, a custom on-chip implementation would likely reduce this power budget. Because the area of this TIA is not reported, we conservatively estimate a footprint of  7,000 $\mu$m$^2$. Inference in the proposed aKAN occurs in $\mathrm{O}(1)$ time, as all computational elements are physically deployed in parallel. The end-to-end inference latency is therefore determined by DAC and ADC conversion times together with the intrinsic RNPU response. We assume that TIA settling and any additional current-to-voltage scaling stages are designed to match the RNPU timescale. Under these conditions, the total inference latency is estimated to be  approximately 600 ns.

For comparison, we consider a 16-bit fixed-point tanh-MLP using the NanGate45 Open Cell Library \cite{FreePDK45}. To enable a controlled architectural comparison, we evaluate the neuron datapath while normalizing across implementations, excluding interconnect and layer-sequencing control that are architecture-dependent.  Each neuron is implemented as a 4-lane single instruction, multiple data (SIMD) unit performing one weight-input multiplication per lane per cycle (four MACs per clock cycle) at moderate hardware cost. The tanh activation is implemented using a 4-lane look-up table evaluated in a single cycle . 

The accelerator is described in Verilog and synthesized using open-source design tools to extract area and power, which are then scaled to the full MLP configuration. In parallel, an equivalent quantized network is trained in PyTorch on the same benchmark task, enabling a direct comparison of approximation error under consistent numerical precision. In this digital architecture, inference latency scales with network depth and width because multiply–accumulate operations must be accumulated sequentially across layers.  For the representative networks considered here, operation at a 500 MHz clock frequency yields an estimated inference latency of approximately 100 ns for MLPs achieving moderate accuracy. 

Figure \ref{fig:hardware}a summarizes the estimated system-level energy per inference as a function of MSE for representative aKAN and MLP configurations (see Supplementary Text 1 for more details). Across moderate-accuracy regimes, aKANs exhibit substantially more favorable energy scaling, reducing the energy per inference by up to three orders of magnitude relative to the digital MLP baseline. Under the assumptions detailed above, the total energy consumption of present aKAN implementations is dominated by mixed-signal peripheral overhead (DACs, ADCs, and TIAs). We pick a representative design with a prediction MSE of $\sim 10^{-3}$, which leads to an estimated energy of approximately 200 pJ per inference. Because this contribution is primarily determined by peripheral circuitry, further reductions in TIA power would proportionally lower the overall inference energy.

By contrast, the digital MLP distributes energy consumption across multiply–accumulate operations, evaluation of activation functions, and the peripheral circuit, as shown in Fig. \ref{fig:hardware}a. The corresponding silicon-area comparison (Fig. \ref{fig:hardware}b) shows that aKAN implementations achieve roughly an order-of-magnitude smaller footprint than their MLP counterparts at MSE $\sim 10^{-3}$, consistent with the parallel, device-level realization of nonlinear computation.

\subsection*{Discussion}
Our results point to a clear message: when learnable nonlinear transformations are treated as primary computational primitives, neural inference can be organized around physical response rather than linear algebra. By realizing nonlinear edge functions directly in matter, aKANs replace the conventional paradigm of linear multiply–accumulate operations combined with digitally emulated nonlinearities with hardware-native nonlinear computation based on tunable device characteristics. This shift defines an alternative design space for edge intelligence, in which efficiency arises not only from accelerating linear operations, but from embedding expressive nonlinear mappings directly within the physical substrate. In contrast to demonstrations focused primarily on device-level nonlinear heterogeneity, the present work emphasizes architectural composability and system-level energy/latency/area estimates including mixed-signal interfacing. 

More broadly, reconfigurable nonlinear devices emerge as modular building blocks for scalable analogue learning systems. Composability, through the use of multiple devices per edge, and structural sparsity, enabled by pruning, provide complementary routes to compactness, robustness, and dynamic resource allocation on reconfigurable arrays. System-level energy analysis further identifies a central architectural constraint: present efficiency is limited primarily by the mixed-signal interface rather than by the RNPU core itself, highlighting the need for joint optimization of devices, circuits, and architectures.

Looking forward, a transition from trained physics toward co-designed physics offers a natural path for progress, in which electrode geometry, biasing, and learning objectives are optimized simultaneously to replace heuristic configuration with principled physical design. Additional gains in energy efficiency are expected from integrated peripheral circuitry and scalable multi-device arrays that remove the need for time-multiplexed operation. Together, these advances outline a realistic route toward compact, low-power edge-inference hardware in which nonlinear computation is native, programmable, and scalable, and suggest a broader paradigm in which information processing is co-designed with the physical properties that implement it.

\section*{Methods} \label{methods}
\subsection*{Experimental setup}
The RNPUs used in this work are boron-doped silicon slabs with eight electrodes placed on top in a circular geometry with a diameter of 300 nm. Six different physical RNPUs were used in this work. Each device is mounted in a measurement enclosure, where seven electrodes are connected to DACs (cDAQ NI-9264, 25 kS/s per channel) and the remaining electrode is connected first to a 2 V/nA gain transimpedance amplifier, enabling indirect measurement of the RNPU output current, and subsequently to an ADC (cDAQ NI-9252, 50 kS/s per channel). 

The RNPU is operated in steady state, i.e., multiple output current samples are acquired after a prudent ramping time of 5 ms for the application of control and input voltages (see Supplementary Fig.~\ref{fig:sup_example}). This delay is dominated by external instrumentation (DAC/ADC and measurement electronics) and does not reflect the intrinsic settling time of the RNPU device itself, which is orders of magnitude faster. This procedure is used both to collect samples for training a surrogate model and for the hardware verification stage.

Due to measurement-setup limitations, only one device can be experimentally verified at a time. Consequently, inference is evaluated in a time-multiplexed fashion, in which the single physical device emulates the different RNPUs in the KAN. The output of each RNPU is measured sequentially by applying the corresponding input and control voltages.

\subsection*{Training and Verification}
All examples are realized using \textit{in-silico} training, i.e., surrogate models of a single or multiple RNPUs are used to train the full KAN off-chip. The surrogate model consists of a seven-input, single output MLP with ReLU activations and five hidden layers containing 90 neurons each. Both aKANs and MLPs are trained using backpropagation with the Adam optimizer, with learning rates ranging from $10^{-3}$ to $10^{-2}$. Initialization includes both random parameter initialization and random selection of the input electrode for each RNPU in the aKAN.

For regression tasks, we use a batch of 1,000 uniformly distributed input samples across the voltage range of the input electrode for training. For classification tasks, we linearly scale the data to the operating range of the RNPU and split the data into 80\% training and 20\% validation sets. The Spirals dataset is generated using a custom script that includes both noise and a configurable number of turns.

The results of the sine example in Fig. \ref{fig:EP_regression}a--b and in Moons and Spirals datasets in Fig. \ref{fig:classification}a are based on experimental hardware measurements, using a single device and a time-multiplexing scheme to evaluate each RNPU device in the aKAN. Due to small mismatches between the surrogate models and real devices, temperature and/or aging effects, we calibrate the aKAN by retraining only the gains after the last layer of RNPUs, using as input the measured outputs of the RNPU facing those gains.

\renewcommand{\refname}{References}
\bibliography{refs.bib}

@article{Krizhevsky2017,
	title = {{ImageNet} classification with deep convolutional neural networks},
	volume = {60},
	number = {6},
	journal = {Communications of the ACM},
	author = {Krizhevsky, Alex and Sutskever, Ilya and Hinton, Geoffrey E.},
	month = may,
	year = {2017},
	pages = {84--90},
}

@article{Vaswani2017,
  title={Attention is all you need},
  author={Vaswani, Ashish and Shazeer, Noam and Parmar, Niki and Uszkoreit, Jakob and Jones, Llion and Gomez, Aidan N and Kaiser, {\L}ukasz and Polosukhin, Illia},
  journal={Advances in neural information processing systems},
  volume={30},
  year={2017}
}

@article{Sozos2026,
author = {Kostas Sozos and Dimitrios Spanos and Stavros Deligiannidis and George Sarantoglou and Nikolaos Passalis and Nikos Pleros and Charis Mesaritakis and Anastasios Tefas and Adonis Bogris},
journal = {Optics Letters},
number = {3},
pages = {664--667},
title = {Photonic Kolmogorov-Arnold networks based on self-phase modulation in nonlinear waveguides},
volume = {51},
month = {Feb},
year = {2026},
}

@article{KAN, 
    title={Kan: Kolmogorov-arnold networks},
    author={Liu, Ziming and Wang, Yixuan and Vaidya, Sachin and Ruehle, Fabian and Halverson, James and Solja{\v{c}}i{\'c}, Marin and Hou, Thomas Y and Tegmark, Max},
    note={arXiv:2404.19756},
    year={2024}
}

@article{MLP_theorem,
  title={Multilayer feedforward networks are universal approximators},
  author={Hornik, Kurt and Stinchcombe, Maxwell and White, Halbert},
  journal={Neural networks},
  volume={2},
  number={5},
  pages={359--366},
  year={1989},
  publisher={Elsevier}
}

@article{Reza2025,
  title={Analogue speech recognition based on physical computing},
  author={Zolfagharinejad, Mohamadreza and Büchel, Julian and Cassola, Lorenzo and Kinge, Sachin and Sarwat Syed, Ghazi and Sebastian, Abu and {van der Wiel}, Wilfred G.},
  journal={Nature},
  volume={645},
  pages={886-892},
  year={2025},
  publisher={Elsevier}
}

@article{activation_function,
  title={Approximation capabilities of multilayer feedforward networks},
  author={Hornik, Kurt},
  journal={Neural networks},
  volume={4},
  number={2},
  pages={251--257},
  year={1991},
  publisher={Elsevier}
}

@article{kan_theorem,
	title = {On the representation of continuous functions of many variables by superposition of continuous functions of one variable and addition},
	volume = {114},
	number = {5},
	journal = {Dokl. Akad. Nauk SSSR},
	author = {Kolmogorov, A.N.},
	year = {1957},
	pages = {935--956},
}

@inproceedings{KAN_digital_hardware_LUTs,
  title={Hardware acceleration of kolmogorov-arnold network (kan) for lightweight edge inference},
  author={Huang, Wei-Hsing and Jia, Jianwei and Kong, Yuyao and Waqar, Faaiq and Wen, Tai-Hao and Chang, Meng-Fan and Yu, Shimeng},
  booktitle={Proceedings of the 30th Asia and South Pacific Design Automation Conference},
  pages={693--699},
  year={2025}
}

@misc{FreePDK45,
    title = {{FreePDK45: 45nm variant from NCSU}},
    url = {https://eda.ncsu.edu/freepdk/freepdk45/},
}

@article{Reza2024,
	title = {Brain-inspired computing systems: a systematic literature review},
	volume = {97},
	number = {6},
	journal = {The European Physical Journal B},
	author = {Zolfagharinejad, Mohamadreza and Alegre-Ibarra, Unai and Chen, Tao and Kinge, Sachin and {van der Wiel}, Wilfred G.},
	month = jun,
	year = {2024},
	pages = {70},
}

@misc{magic,
	title = {{MAGIC} {Gamma} {Telescope}},
	doi = {10.24432/C52C8B},
	publisher = {UCI Machine Learning Repository},
	author = {{R. Bock}},
	year = {2004},
}

@article{codrna,
	title = {Detection of non-coding {RNAs} on the basis of predicted secondary structure formation free energy change},
	volume = {7},
	number = {1},
	journal = {BMC Bioinformatics},
	author = {Uzilov, Andrew V and Keegan, Joshua M and Mathews, David H},
	month = dec,
	year = {2006},
	pages = {173},
}

@misc{skin,
	title = {Skin {Segmentation}},
	doi = {10.24432/C5T30C},
	publisher = {UCI Machine Learning Repository},
	author = {Rajen Bhatt, Abhinav Dhall},
	year = {2009},
}

@article{Chen2020,
	title = {Classification with a disordered dopant-atom network in silicon},
	volume = {577},
	number = {7790},
	journal = {Nature},
	author = {Chen, Tao and {van Gelder}, Jeroen and {van de Ven}, Bram and Amitonov, Sergey V. and {de Wilde}, Bram and {{H.-C.} Ruiz Euler} and Broersma, Hajo J. and Bobbert, Peter A. and Zwanenburg, Floris A. and {van der Wiel}, Wilfred G.},
	month = jan,
	year = {2020},
	pages = {341--345},
}

@article{Chen2021,
	title = {1/ \textit{f} {Noise} and {Machine} {Intelligence} in a {Nonlinear} {Dopant} {Atom} {Network}},
	volume = {1},
	number = {3},
	journal = {Small Science},
	author = {Chen, Tao and Bobbert, Peter A. and {van der Wiel}, Wilfred G.},
	month = mar,
	year = {2021},
	pages = {2170006},
}

@article{RuizEuler2020,
	title = {A deep-learning approach to realizing functionality in nanoelectronic devices},
	volume = {15},
	number = {12},
	journal = {Nature Nanotechnology},
	author = {{{H.-C.} Ruiz Euler} and Boon, Marcus N. and Wildeboer, Jochem T. and {van de Ven}, Bram and Chen, Tao and Broersma, H. J. and Bobbert, Peter A. and {van der Wiel}, Wilfred G.},
	month = dec,
	year = {2020},
	pages = {992--998},
}

@article{RuizEuler2021,
	title = {Dopant network processing units: towards efficient neural network emulators with high-capacity nanoelectronic nodes},
	volume = {1},
	issn = {2634-4386},
	shorttitle = {Dopant network processing units},
	number = {2},
	journal = {Neuromorphic Computing and Engineering},
	author = {{{H.-C.} Ruiz Euler} and Alegre-Ibarra, Unai and {van de Ven}, Bram and Broersma, H. J. and Bobbert, Peter A and {van der Wiel}, Wilfred G},
	month = dec,
	year = {2021},
	pages = {024002},
}

@article{Ven2023,
	title = {Dopant network processing units as tuneable extreme learning machines},
	volume = {5},
	issn = {2673-3013},
	journal = {Frontiers in Nanotechnology},
	author = {{van de Ven}, B. and Alegre-Ibarra, U. and Lemieszczuk, P. J. and Bobbert, P. A. and Ruiz-Euler, Hans-Christian and {van der Wiel}, W. G.},
	month = mar,
	year = {2023},
	pages = {1055527},
}

@article{Boon2025,
	title = {Gradient descent in materia through homodyne gradient extraction},
	volume = {16},
	number = {1},
	journal = {Nature Communications},
	author = {Boon, Marcus N. and Cassola, Lorenzo and {{H.-C.} Ruiz Euler} and Chen, Tao and {van de Ven}, Bram and Alegre-Ibarra, Unai and Bobbert, Peter A. and {van der Wiel}, Wilfred G.},
	month = nov,
	year = {2025},
	pages = {10272},
}

@article{George2016,
	title = {A {Programmable} and {Configurable} {Mixed}-{Mode} {FPAA} {SoC}},
	journal = {IEEE Transactions on Very Large Scale Integration (VLSI) Systems},
	author = {George, Suma and Kim, Sihwan and Shah, Sahil and Hasler, Jennifer and Collins, Michelle and Adil, Farhan and Wunderlich, Richard and Nease, Stephen and Ramakrishnan, Shubha},
	year = {2016},
	pages = {1--9},
}

@misc{moons,
	title = {Moons dataset},
	url = {https://scikit-learn.org/stable/modules/generated/sklearn.datasets.make_moons.html},
}

@misc{spirals,
	title = {Spirals dataset},
	url = {https://conx.readthedocs.io/en/latest/Two-Spirals.html},
}

@incollection{RezaThesis2025,
	address = {Enschede, The Netherlands},
	title = {Chapter 6: {Towards} nonlinear function approximation with {RNPUs}},
	isbn = {978-90-365-6810-4},
	doi = {10.3990/1.9789036568104},
	booktitle = {Information processing with silicon-based nonlinear computing units},
	publisher = {University of Twente},
	author = {Zolfagharinejad, Mohamadreza},
	month = aug,
	year = {2025},
}

@article{AlegreIbarra2023, year = {2023}, publisher = {The Open Journal}, volume = {8}, number = {90}, pages = {5573}, author = {Alegre-Ibarra, Unai and {{H.-C.} Ruiz Euler} and A.Mollah, Humaid and Petrov, Bozhidar P. and Sastry, Srikumar S. and Boon, Marcus N. and de Jong, Michel P. and Zolfagharinejad, Mohamadreza and Uitzetter, Florentina M. j. and van de Ven, Bram and de Almeida, António J. Sousa and Kinge, Sachin and van der Wiel, Wilfred G.}, title = {brains-py, A framework to support research on energy-efficient unconventional hardware for machine learning}, journal = {Journal of Open Source Software} }

@conference{DAC,
	address = {Bangkok, Thailand},
	title = {Design of {Relaxation} {Digital}-to-{Analog} {Converters} for {Internet} of {Things} {Applications} in 40nm {CMOS}},
	booktitle = {2019 {IEEE} {Asia} {Pacific} {Conference} on {Circuits} and {Systems} ({APCCAS})},
	publisher = {IEEE},
	author = {Rubino, Roberto and Crovetti, Paolo S. and Aiello, Orazio},
	month = nov,
	year = {2019},
	pages = {13--16},
}

@article{ADC,
	title = {A 0.9-{V} 12-bit 100-{MS}/s 14.6-{fJ}/{Conversion}-{Step} {SAR} {ADC} in 40-nm {CMOS}},
	volume = {26},
	issn = {1063-8210, 1557-9999},
	number = {10},
	journal = {IEEE Transactions on Very Large Scale Integration (VLSI) Systems},
	author = {Luo, Jian and Li, Jing and Ning, Ning and Liu, Yang and Yu, Qi},
	month = oct,
	year = {2018},
	pages = {1980--1988},
}

@article{Yu2024,
      title={KAN or MLP: A Fairer Comparison}, 
      author={Runpeng Yu and Weihao Yu and Xinchao Wang},
      year={2024},
      note={arXiv:2407.16674},
      archivePrefix={arXiv},
}

@article{Kaspar2021,
	title = {The rise of intelligent matter},
	volume = {594},
	number = {7863},
	journal = {Nature},
	author = {Kaspar, C. and Ravoo, B. J. and {van der Wiel}, W. G. and Wegner, S. V. and Pernice, W. H. P.},
	year = {2021},
	pages = {345--355},
}

@article{Jaeger2023,
	title = {Toward a formal theory for computing machines made out of whatever physics offers},
	volume = {14},
	language = {en},
	number = {1},
	journal = {Nature Communications},
	author = {Jaeger, Herbert and Noheda, Beatriz and {van der Wiel}, Wilfred G.},
	month = aug,
	year = {2023},
	pages = {4911},
}

@article{Taglietti2026,
	title = {Learning {Nonlinear} {Heterogeneity} in {Physical} {Kolmogorov}-{Arnold} {Networks}},
	note = {arXiv:2601.15340},
	author = {Taglietti, Fabiana and Pulici, Andrea and Roxburgh, Maxwell and Seguini, Gabriele and Vidamour, Ian and Menzel, Stephan and Franco, Edoardo and Laus, Michele and Vasilaki, Eleni and Perego, Michele and Hayward, Thomas J. and Fanciulli, Marco and Gartside, Jack C.},
	year = {2026},
}

@article{Hoang2025,
    author = {Hoang, Duc and Gupta, Aarush and Harris, Philip},
    title = {KANEL{\'E}: Kolmogorov-Arnold Networks for Efficient LUT-based Evaluation},
    note = {arXiv:2512.12850},
}

@article{Zhang2021,
	title = {A {Survey} on {Neural} {Network} {Interpretability}},
	volume = {5},
	number = {5},
	journal = {IEEE Transactions on Emerging Topics in Computational Intelligence},
	author = {Zhang, Yu and Tino, Peter and Leonardis, Ales and Tang, Ke},
	month = oct,
	year = {2021},
	pages = {726--742},
}

@article{Thompson2020,
	title = {The {Computational} {Limits} of {Deep} {Learning}},
	author = {Thompson, Neil C. and Greenewald, Kristjan and Lee, Keeheon and Manso, Gabriel F.},
	year = {2020},
	note = {Arxiv.2007.05558},
}

@article{PhotonicKAN, 
    title={Photonic KAN: a Kolmogorov-Arnold network inspired efficient photonic neuromorphic architecture},
    author={Peng, Yiwei and Hooten, Sean and Yu, Xinling and Van Vaerenbergh, Thomas and Yuan, Yuan and Xiao, Xian and Tossoun, Bassem and Cheung, Stanley and Fiorentino, Marco and Beausoleil, Raymond},
    note={arXiv:2408.08407},
    year={2024}
}

@article{Sudarshan2025,
    author = {Sudarshan, Chirag and Manea, Paul and Strachan, J. P.},
    title={A Kolmogorov–Arnold Compute-in-Memory (KA-CIM) Hardware Accelerator with High Energy Efficiency and Flexibility},
    note = {Reseach Square preprint},
    year={2025}}

@article{Li2025,
  title={Fully analogue in-memory neural computing via quantum tunneling effect},
  author={Li, Songyuan and Wang, Teng and Tang, Jinrong and Liu, Ruiqi and Li, Haoyu and Lu, Yuyao and Xu, Feng and Gao, Bin and Xie, Can and Zhu, Xiangwei},
  journal={arXiv:2510.23638},
  year={2025}
}

@article{Sutton2019,
  title={The bitter lesson},
  author={Sutton, Richard},
  journal={Incomplete Ideas (blog)},
  year={2019},
  url={http://www.incompleteideas.net/IncIdeas/BitterLesson.html}
}

@article{Hooker2021,
  title={The hardware lottery},
  author={Hooker, Sara},
  journal={Communications of the ACM},
  volume={64},
  number={12},
  pages={58--65},
  year={2021},
  publisher={ACM New York, NY, USA}
}

@article{Laydevant2024,
  title={The hardware is the software},
  author={Laydevant, J{\'e}r{\'e}mie and Wright, Logan G and Wang, Tianyu and McMahon, Peter L},
  journal={Neuron},
  volume={112},
  number={2},
  pages={180--183},
  year={2024},
  publisher={Elsevier}
}

@article{Markovic2020,
  title={Physics for neuromorphic computing},
  author={Markovi{\'c}, Danijela and Mizrahi, Alice and Querlioz, Damien and Grollier, Julie},
  journal={Nature Reviews Physics},
  volume={2},
  number={9},
  pages={499--510},
  year={2020},
}

@article{Singh2025,
  title={The design of analogue in-memory computing tiles},
  author={Singh, Abhairaj and Le Gallo, Manuel and Vasilopoulos, Athanasios and Luquin, Jose and Narayanan, Pritish and Burr, Geoffrey W and Sebastian, Abu},
  journal={Nature Electronics},
  pages={1156--1169},
  volume={8},
  year={2025},
}

@article{Mehonic2022,
  title={Brain-inspired computing needs a master plan},
  author={Mehonic, Adnan and Kenyon, Anthony J},
  journal={Nature},
  volume={604},
  number={7905},
  pages={255--260},
  year={2022},
}

@article{Kasai2022,
  title={Semiconductor technologies and related topics for implementation of electronic reservoir computing systems},
  author={Kasai, Seiya},
  journal={Semiconductor Science and Technology},
  volume={37},
  number={10},
  pages={103001},
  year={2022},
  publisher={IOP Publishing}
}

@article{Liang2024,
  title={Physical reservoir computing with emerging electronics},
  author={Liang, Xiangpeng and Tang, Jianshi and Zhong, Yanan and Gao, Bin and Qian, He and Wu, Huaqiang},
  journal={Nature Electronics},
  volume={7},
  number={3},
  pages={193--206},
  year={2024},
  publisher={Nature Publishing Group UK London}
}

@article{Momeni2025,
  title={Training of physical neural networks},
  author={Momeni, Ali and Rahmani, Babak and Scellier, Benjamin and Wright, Logan G and McMahon, Peter L and Wanjura, Clara C and Li, Yuhang and Skalli, Anas and Berloff, Natalia G and Onodera, Tatsuhiro and others},
  journal={Nature},
  volume={645},
  number={8079},
  pages={53--61},
  year={2025},
  publisher={Nature Publishing Group UK London}
}

@article{Mathew2022,
  title={Low input-resistance low-power transimpedance amplifier design for biomedical applications},
  author={Mathew, M and Hart, BL and Hayatleh, Khaled},
  journal={Analog Integrated Circuits and Signal Processing},
  volume={110},
  number={3},
  pages={527--534},
  year={2022},
  publisher={Springer}
}

\subsection*{Acknowledgments} \label{acknowledgments}
We thank M. H. Siekman, J. G. M. Sanderink and M. Schremb for technical support, U. Alegre-Ibarra, A. J. Annema, P. A. Bobbert, L. Cassola, T. Chen, R. J. C. Cool and J. Kareem for stimulating discussions.
We acknowledge financial support from Toyota Motor Europe N.V., the IMAGINE project funded by the Dutch Research Council (NWO) KIC grant no. KICH1.ST04.22.033, the HYBRAIN project funded by the European Union’s Horizon Europe research and innovation programme under grant agreement no. 101046878. This work was further funded by the Deutsche Forschungsgemeinschaft (DFG, German Research Foundation) grant no. SFB 1459/2 2025–433682494.

\subsection*{Author contributions} \label{contributions}
M.Z. and W.G.v.d.W. conceived the project. M.E., M.Z., and S.v.d.B. carried out the experiments. All authors contributed to data analysis and interpretation. All authors contributed to the writing of the manuscript and to revisions. W.G.v.d.W. supervised the project. 

\subsection*{Competing interests}
The authors declare that they have no competing interests.

\subsection*{Data availability}
The data that support the findings of this study are available from the corresponding author upon reasonable request.

%%%%%%%%%%%%%%%% END OF MAIN TEXT %%%%%%%%%%%%%%%

\newpage

%%%%%%%%%%%%%%%% START OF SUPPLEMENT %%%%%%%%%%%%%%%

% Figures, tables, equations and pages in the supplement are numbered S1, S2 etc.
\renewcommand{\figurename}{Supplementary Figure}
\renewcommand{\tablename}{Supplementary Table}
\setcounter{figure}{0}
\setcounter{table}{0}
\setcounter{equation}{0}
\setcounter{page}{1} % not 0 as \newpage already started a supplementary page
% References continue the numbering from the main text.

%%%%%%%%%%%%%%%% SUPPLEMENT TITLE PAGE %%%%%%%%%%%%%%%

\begin{center}
\section*{Supplementary Information for\\}
\textbf{Physical Analogue Kolmogorov-Arnold Networks based on Reconfigurable Nonlinear-Processing Units}

Manuel~Escudero$^{1}$,
Mohamadreza~Zolfagharinejad$^{1}$,
Sjoerd~van~den~Belt$^{2}$,\\
Nikolaos~Alachiotis$^{2}$,
Wilfred~G.~van~der~Wiel$^{1,3\ast}$\\

\small$^{1}$MESA+ Institute and BRAINS Center for Brain-Inspired Computing, \and\ 
\small University of Twente, Enschede, The Netherlands.\\
\small$^{2}$CAES Group and BRAINS Center for Brain-Inspired Computing, \and\
\small University of Twente, Enschede, The Netherlands. \\
\small$^{3}$Institute of Physics, University of M\"{u}nster, M\"{u}nster, Germany.

\small$^\ast$Corresponding author. Email: w.g.vanderwiel@utwente.nl\\
\end{center}

\subsubsection*{This PDF file includes:}

Supplementary Text 1\\
Supplementary Figures 1-2\\
Supplementary Tables 1-2\\

\newpage

\subsubsection*{Supplementary Text 1. Estimation of the energy consumption}

The total estimated energy of an aKAN of $n_{I}$ inputs, $n_{O}$ outputs, $L$ hidden layers with $W$ nodes per hidden layer, and $n_{C}$ control voltages, is expressed as the sum of contributions of the main building blocks: 
\begin{equation}
    E_{\mathrm{TOTAL}} = E_{\mathrm{ADC}} + E_{\mathrm{DAC}} + E_{\mathrm{TIA}} + E_{\mathrm{RNPU}}
\end{equation}
For the evaluation of a function at $P$ sample points, we explicitly count the conversions performed by the DACs and the ADC. The DACs dedicated to control voltages set the voltage values once, while the DACs for inputs perform conversions for each sample. Therefore, the total energy consumed by the DACs during the function evaluation is 
\begin{equation}
E_{\mathrm{DAC}} = n_{\mathrm{I}} E_{\mathrm{ACONV}} + n_{\mathrm{C}}E_{\mathrm{CONV}}
\end{equation}
where $E_{\mathrm{ACONV}}$ is the energy per conversion of the selected DAC. The ADCs convert $n_{\mathrm{O}}$ outputs (in this work, $n_{\mathrm{O}} = 1$) each inference at $E_{\mathrm{DCONV}}$ energy per conversion, which yields $E_{\mathrm{ADC}} = n_{\mathrm{O}} E_{\mathrm{DCONV}}$.

The remaining energy contributions are computed from their estimated power and the inference latency $t_{\mathrm{d}}$. More concretely, ${E_{\mathrm{TIA}} = WL P_{\mathrm{TIA}} t_{\mathrm{d}}}$ and ${E_{\mathrm{RNPU}} = N_{\mathrm{RNPU}} P_{\mathrm{RNPU}} t_{\mathrm{d}}}$, accounting for the number of nodes $N_{\mathrm{NODES}}$ and the number of RNPUs $N_{\mathrm{RNPU}}$ in the aKAN. The energy consumed by the programmable resistors (memristors) is neglected relative to the TIA consumption. The inference latency is the sum of DAC, RNPU layers, and ADC latencies, i.e., $t_{\mathrm{d}} = t_{\mathrm{DAC}} +  t_{\mathrm{RNPU}} L + t_{\mathrm{ADC}}$.

\newpage

% %%%%%%%%%%%%%%%% SUPPLEMENTARY FIGURES %%%%%%%%%%%%%%%

\begin{figure} % Do not use \begin{figure*}
	\centering
     \makebox[\textwidth][c]{
     \includegraphics[width=18cm]{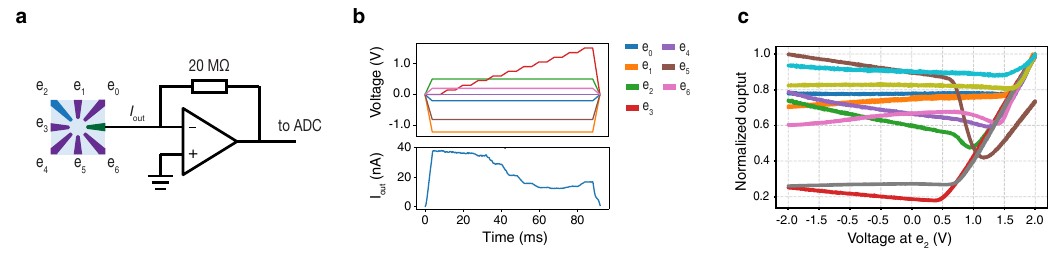}
     }
	\caption{\textbf{RNPU experimental evaluation.} \textbf{a}, Circuit scheme, including the input and/or control electrodes $e_0$ to $e_6$ connected to the DACs and the output electrode connected to the transimpedance amplifier, which provides a voltage to the ADC proportional to the RNPU output current. \textbf{b}, Top: applied control (constant) and input (variable) voltages. Bottom: corresponding raw output. For each input evaluation, the device first undergoes a settling (ramping) period to reach a steady output. This is followed by one or more output samplings. These samples are then averaged to produce a single output value that represents the result of that evaluation (not shown). \textbf{c}, Example of 10 RNPU output currents while using electrode $e_2$ as input electrode and setting the rest of the electrodes to random control voltages. The output levels are already averaged and settling periods are discarded. The output curves have been normalized.}
	\label{fig:sup_example}
\end{figure}

\begin{figure} % Do not use \begin{figure*}
	\centering
	\includegraphics[]{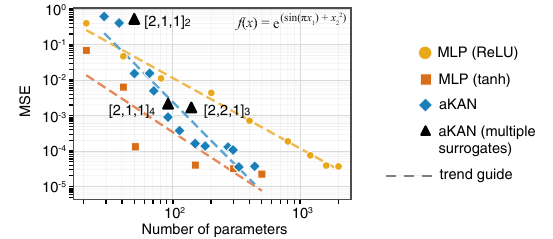} 
	\caption{\textbf{Scalability of aKANs mixing surrogate models of different RNPU devices.} Extension of Fig. \ref{fig:regression_scaling}b including simulations of aKANs with up to six different surrogate models (black triangles).}
	\label{figS:mult_devices_regression}
\end{figure}

% %%%%%%%%%%%%%%%% SUPPLEMENTARY TABLES %%%%%%%%%%%%%%%

\begin{table} % Do not use \begin{table*}
	\centering
	% Captions go above tables
	\caption{\textbf{Simulated MLP configurations for the function approximation task.}}
	\label{tab:MLP_conf} % give each table a logical label name

	\begin{tabular}{lccr} % four columns, alignment for each
		\\
		\hline
		  Function & Networks \\
		\hline
		Fig. \ref{fig:regression_scaling}a  & [1,50,1], [1,100,1], [1,150,1], [1,200,1], [1,250,1], [1,300,1], [1,350,1], [1,400,1] \\
		Fig. \ref{fig:regression_scaling}b & [2,5,1], [2,10,1], [2,20,1], [2,50,1], [2,100,1], [2,200,1], [2,300,1], [2,400,1], [2,500,1] \\
		Fig. \ref{fig:regression_scaling}c & [4,50,1], [4,100,1],[4,150,1],[4,200,1],[4,250,1],[4,300,1],[4,10,10,1], \\ 
        & [4,15,15,1], [4,20,20,1],[4,50,50,1],[4,200,200,1],[4,300,300,1],[4,400,400,1] \\

		\hline
	\end{tabular}
\end{table}

\begin{table} % Do not use \begin{table*}
	\centering
	% Captions go above tables
	\caption{\textbf{Simulated aKAN configurations for the function approximation task.}}
	\label{tab:aKAN_conf} % give each table a logical label name

	\begin{tabular}{lccr}
		\\
		\hline
		  Function & Networks & RNPUs per EP\\
		\hline
		Fig. \ref{fig:regression_scaling}a  & [1,1] & 5, 10, 15, 20, 25\\
         &  & 30, 35, 40, 45, 50\\
		Fig. \ref{fig:regression_scaling}b & [2,1,1], [2,2,1], [2,3,1], [2,4,1], [2,5,1] & 1, 2, 3, 4, 5\\
         & [2,1,1,1], [2,2,2,1], [2,3,3,1], [2,4,4,1], [2,5,5,1] & \\
		Fig. \ref{fig:regression_scaling}c & [4,2,1], [4,3,1], [4,4,1], [4,1,1,1], [4,2,2,1], [4,3,3,1] & 1, 2, 3\\
         & [4,4,4,1], [4,1,1,1,1], [4,2,2,2,1], [4,3,3,3,1], [4,4,4,4,1] & \\

		\hline
	\end{tabular}
\end{table}

\end{document}